\documentclass[longbibliography, nature, amsmath, amssymb, amsfonts, twocolumn,superscriptaddress, footinbib]{revtex4-1}
\usepackage[german,american,english]{babel}
\usepackage{graphics}
\usepackage{dcolumn}
\usepackage{bm}
\usepackage{amssymb}
\usepackage{amsmath}
\usepackage{amsfonts}
\usepackage{epsfig}
\usepackage{mathbbol}
\usepackage{latexsym}
\usepackage{dcolumn}
\usepackage{subfigure}
\usepackage{hyperref}
\usepackage{float}
\usepackage[normalem]{ulem}
\usepackage[usenames, dvipsnames]{color}
\usepackage{epstopdf}
\usepackage{tabularx}
\usepackage[ruled,vlined]{algorithm2e}
\usepackage{colortbl}
\usepackage{diagbox}

\hypersetup{                    
	colorlinks,
	citecolor=blue,
	filecolor=blue,
	linkcolor=blue,
	urlcolor=blue
}
\usepackage[utf8]{inputenc}

\begin{document}
	\title{Quantum imaginary time evolution steered by reinforcement learning}

	\author{Chenfeng Cao}\thanks{These authors contributed equally.}
	\affiliation{Department of Physics, The Hong Kong University of Science and Technology, Hong Kong, China}
	
	\author{Zheng An} \thanks{These authors contributed equally.}
	\affiliation{Institute of Physics, Beijing National Laboratory for Condensed Matter Physics, Chinese Academy of Sciences, Beijing 100190, China}
	\affiliation{School of Physical Sciences, University of Chinese Academy of Sciences, Beijing 100049, China}
	
	\author{Shi-Yao Hou}
	\affiliation{Department of Physics, The Hong Kong University of Science and Technology, Hong Kong, China}
	
	\author{D. L. Zhou}
	\email{zhoudl72@iphy.ac.cn}
	\affiliation{Institute of Physics, Beijing National Laboratory for Condensed Matter Physics, Chinese Academy of Sciences, Beijing 100190, China}
	\affiliation{School of Physical Sciences, University of Chinese Academy of Sciences, Beijing 100049, China}
	\affiliation{Collaborative Innovation Center of Quantum Matter, Beijing 100190, China}
	\affiliation{Songshan Lake Materials Laboratory, Dongguan, Guangdong 523808, China}

	\author{Bei Zeng}%
	\email{zengb@ust.hk}
	\affiliation{Department of Physics, The Hong Kong University of Science and Technology, Hong Kong, China}
	
	\begin{abstract}
		The quantum imaginary time evolution is a powerful algorithm for preparing the ground and thermal states on near-term quantum devices. However, algorithmic errors induced by Trotterization and local approximation severely hinder its performance. Here we propose a deep reinforcement learning-based method to steer the evolution and mitigate these errors. In our scheme, the well-trained agent can find the subtle evolution path where most algorithmic errors cancel out, enhancing the fidelity significantly. We verified the method's validity with the transverse-field Ising model and the Sherrington-Kirkpatrick model. Numerical calculations and experiments on a nuclear magnetic resonance quantum computer illustrate the efficacy. The philosophy of our method, eliminating errors with errors, sheds light on error reduction on near-term quantum devices.
	\end{abstract}

	\date{\today}
	\maketitle

	\section{Introduction}
	Quantum computers promise to solve some computational problems much faster than classical computers in the future. However, large-scale fault-tolerant quantum computers are still years away. On noisy intermediate-scale quantum (NISQ) devices~\cite{preskill2018quantum}, quantum noise strictly limits the depth of reliable circuits, which makes many quantum algorithms unrealistic, e.g., Shor algorithm for factorization~\cite{shor1999polynomial}, Harrow-Hassidim-Lloyd algorithm for solving linear systems of equations~\cite{harrow2009quantum}. Nevertheless, there exist quantum algorithms that are well suited for NISQ devices and may achieve quantum advantage with practical applications, such as the variational quantum algorithms~\cite{kandala2017hardware, farhi2014quantum, cerezo2020variational2, bharti2021noisy, romero2017quantum, bondarenko2020quantum, cao2021noise, wiersema2020exploring, cao2021energy}, the quantum imaginary time evolution~\cite{mcardle2019variational,motta2020determining, yeter2020practical}, quantum annealing~\cite{albash2018adiabatic}.

	The quantum imaginary time evolution (QITE) is a promising near-term algorithm to find the ground state of a given Hamiltonian. It has also been applied to prepare thermal states, simulate open quantum systems, and calculate finite temperature properties~\cite{zeng2020variational, kamakari2021digital, sun2021quantum}. A pure quantum state is said to be $k$-UGS if it is the unique ground state of a $k$-local Hamiltonian $\hat{H}=\sum_{j=1}^m \hat{h}[j]$, where each local term $\hat{h}[j]$ acts on at most $k$ neighboring qubits. Any $k$-UGS state can be uniquely determined by its $k$-local reduced density matrices among pure states (which is called $k$-UDP) or even among all states (which is called $k$-UDA)~\cite{aharonov2018quantum,chen2012ground}. The QITE algorithm is well suited for preparing $k$-UGS states with a relatively small $k$. We start from an initial state $|\Psi_{\mathrm{init}}\rangle$, which is non-orthogonal to the ground state of the target Hamiltonian. The final state after long-time imaginary time evolution
	\begin{equation}
		\lim _{\beta \rightarrow \infty} e^{-\beta \hat{H}}\left|\Psi_{\mathrm{init}}\right\rangle
	\end{equation}
	has very high fidelity with the $k$-UGS state. If the ground state of $\hat{H}$ is degenerate, the final state still falls into the ground state space. Trotter-Suzuki decomposition can simulate the evolution,
	
	\begin{equation}\label{trotter}
		e^{-\beta \hat{H}}=\left(e^{-\Delta \tau \hat{h}[1]} e^{-\Delta \tau \hat{h}[2]} \ldots e^{-\Delta \tau \hat{h}[m]}\right)^{n}+\mathcal{O}(\Delta \tau ^2),
	\end{equation}
	where $\Delta \tau$ is the step interval, $n=\frac{\beta}{\Delta \tau}$ is the number of Trotter steps. Trotter error subsumes terms of order $\Delta \tau ^2$ and higher. On NISQ devices, Trotter error is difficult to reduce due to the circuit depth limits and Trotterized simulation cannot be implemented accurately~\cite{smith2019simulating}.

	Since we can only implement unitary operations on a quantum computer, the main idea of the QITE is to replace each non-unitary step $e^{-\Delta \tau \hat{h}[j]}$ by a unitary evolution $e^{-i\Delta \tau \hat{A}[j]}$ such that \begin{equation}\label{local_approx}
		|\bar{\Psi}'\rangle = \frac{e^{-\Delta \tau \hat{h}[j]}}{\sqrt{\langle\Psi|e^{-2 \Delta \tau \hat{h}[j]}| \Psi\rangle}}|\Psi\rangle \approx e^{-i \Delta \tau \hat{A}[j]}|\Psi\rangle,
	\end{equation}
	where $|\Psi\rangle$ is the state before this step. $\hat{A}[j]$ acts on $D$ neighboring qubits and can be determined by measurements on $|\Psi\rangle$. For details of the local approximation see Appendix~\ref{sec1}. If the domain size $D$ equals the system size $N$, there always exists $\hat{A}[j]$, such that the approximation sign of Eq.~\eqref{local_approx} becomes an equal sign. However, exp($D$) local gates are required to implement a generic $D$-qubit unitary, and we also need to measure exp($D$) observables to determine $\hat{A}[j]$. The exponential resource of measurements and computation makes a large domain size $D$ unfeasible, and we can only use a small one on real devices. This brings the local approximation (LA) error.
	
	Trotter error and the LA error are two daunting challenges in the QITE. These algorithmic errors accumulate with the increase of steps $n$,  which severely weakens the practicability of the QITE. On NISQ computers, a circuit with too many noisy gates is unreliable, and the final measurements give no helpful information. Therefore we cannot use a small step interval $\Delta \tau$ to reduce Trotter errors since this would increase the circuit depth, and noise would dominate the final state. The number of Trotter steps is a tradeoff between quantum noise and Trotter error. For the QITE with large-size systems, we need more Trotter steps and larger domain sizes, which seems hopeless on current devices. There exist some techniques to alleviate the problem, Refs.~\cite{nishi2021implementation, gomes2020efficient, gome2021adaptive} illustrated some variants of the QITE algorithm with shallower circuits. Refs.~\cite{sun2021quantum, ville2021leveraging} used Hamiltonian symmetries, error mitigation, and randomized compiling to reduce the required quantum resources and improve the fidelity. 

	Reinforcement learning (RL) is an area of machine learning concerned with how intelligent agents interact with an environment to perform well in a specific task. It achieved great success in classical
	games~\cite{alphago,alphagozero,alphazero,atari,alphastar, agostinelli2019solving}, and has been employed in quantum computing problems, such as quantum control~\cite{bukov2018reinforcement, niu2019universal, zhang2019does, gate, ladder, yao2020noise}, quantum circuit optimization~\cite{khairy2020learning, fosel2021quantum, ostaszewski2021reinforcement}, the quantum approximate optimization~\cite{wauters2020reinforcement, yao2021reinforcement}, and quantum annealing~\cite{lin2020quantum}. Quantum computing, in turn, can enhance classical reinforcement learning~\cite{jerbi2021quantum, saggio2021experimental}.
	
	In this work, we propose a deep reinforcement learning-based method to steer the QITE and mitigate algorithmic errors. In our method, we regard the ordering of local terms in the QITE as the environment and train an intelligent agent to take actions (i.e., exchange adjacent terms) to minimize the final state energy. RL is well suited for this task since the state and action space can be pretty large. We verified the validity of our method with the transverse-field Ising model and the Sherrington-Kirkpatrick model. The RL agent can mitigate most algorithmic errors and decrease the final state energy. Our work pushes the QITE algorithm to more practical applications in the NISQ era.

	\section{Methods}
	The RL process is essentially a finite Markov decision process~\cite{sutton}. This process is described as a state-action-reward sequence, a state $s_t$ at time $t$ is transmitted into a new state $s_{t+1}$ together with giving a scalar reward $R_{t+1}$ at time $t+1$ by the action $a_t$ with the transmission probability $p(s_{t+1};R_{t+1}|s_t;a_t)$. In a finite Markov decision process, the state set, the action set and the reward set are finite. The total discounted return at time $t$ is
	\begin{equation}
		G_{t}=\sum_{k=0}^{\infty}\gamma^{k}R_{t+k+1},
	\end{equation}
	where $\gamma$ is the discount rate and $0\le\gamma\le1$. 
	
	The goal of RL is to maximize the total discounted return for each state and action selected by the policy $\pi$, which is specified by a conditional probability of action $a$ for each state $s$, denoted as $\pi(a|s)$.

	In this work, we use distributed proximal policy optimization (DPPO)~\cite{DPPO1,DPPO2}, a model-free reinforcement learning algorithm with the actor-critic architecture. The agent has several distributed evaluators, and each evaluator consists of two components: an actor-network that computes a policy $\pi$, according to which the actions are probabilistically chosen; a critic-network that computes the state value $V(s)$, which is an estimate of the total discounted return from state $s$ and the following policy $\pi$. Using multiple evaluators can break the unwanted correlations between data samples and make the training process more stable. The RL agent updates the neural network weights synchronously. For more details of DPPO see Appendix~\ref{sec2}.
	
	\begin{figure}[t]
		\includegraphics[width=9cm,keepaspectratio]{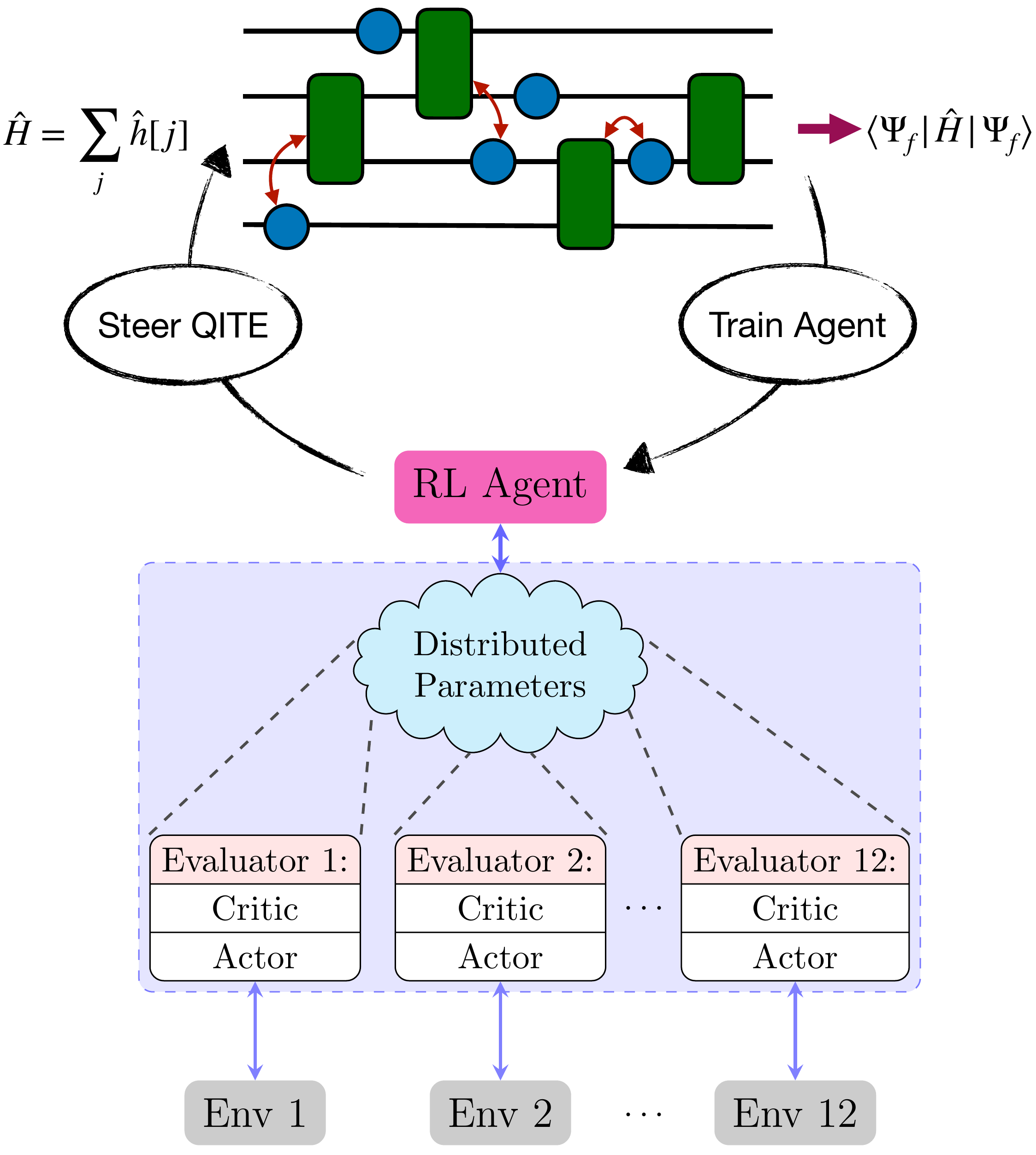}
		\caption{\textbf{Schematic of steering the quantum imaginary time evolution with reinforcement learning.} The colored symbols represent single-qubit (blue dots) and two-qubit (green rectangles) non-unitary operations $\{e^{-\Delta \tau \hat{h}[j]}\}$. The reinforcement learning (RL) agent, realized by neural networks, interacts with different environments (Env 1, Env 2, \dots, Env 12) and optimizes the operation ordering to minimize the output state energy.}
		\label{diagram}
	\end{figure}

\begin{figure*}[t]
	\centerline{\includegraphics[width=18cm]{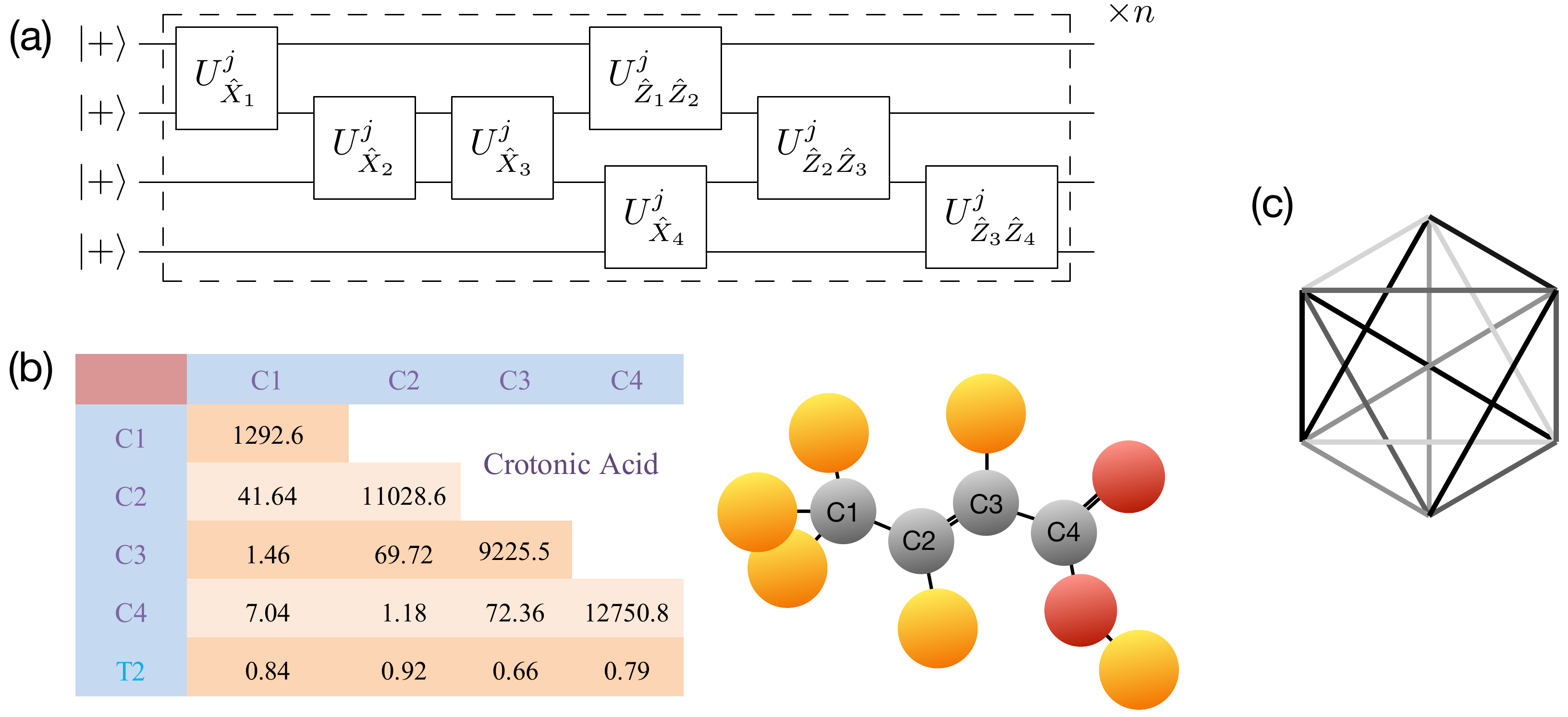}}
	\caption{\textbf{Theoretical and experimental setup.} (a) Quantum circuit of the quantum imaginary time evolution (QITE) for the transverse-field Ising model with $4$ qubits and $n$ Trotter steps. $U_{\hat{P}}^j$ represents the unitary operator that approximates $e^{-\Delta \tau \hat{P}}$ for the local Hamiltonian $\hat{P}$ in the $j$-th Trotter step. (b) Molecule structure and nuclei parameters of the nuclear magnetic resonance processor. The molecule has 4 Carbon atoms C1, C2, C3, and C4. Diagonal entries of the table are the chemical shifts in $\text{Hz}$, off-diagonal entries of the table are the $J$-couplings between two corresponding nuclei. The $T_2$ row gives the relaxation time of each nucleus. (c) A six-vertex complete graph with weighted edges. Different shades of grey represent different couplings in the Sherrington-Kirkpatrick model.}
	\label{fig:2}
\end{figure*}

	The objective of the agent is to maximize the cumulative reward under a parameterized policy $\pi_{\boldsymbol{\theta}}$:
	\begin{equation}
		J\left(\pi_{\boldsymbol{\theta}}\right) = \mathbb{E}_{\pi_{\boldsymbol{\theta}}}\left[\sum_{t=0}^{\infty} \gamma^{t} R\left(s_{t}\right)\right].
		\label{eq:pg}
	\end{equation}\\
	In our task, the environment state is the ordering of local terms in each Trotter step, and the state space size is $(m!)^n$. The agent observes the full state space at the learning stage, i.e., we deal with a fully observed Markov process. We define the action set by whether or not to exchange two adjacent operations in Eq.~\eqref{trotter}. Note that even if two local terms $\hat{h}[j]$ and $\hat{h}[j+1]$ commute, their unitary approximations $e^{-i \Delta \tau \hat{A}[j]}$ and $e^{-i \Delta \tau \hat{A}[j+1]}$ are state-dependent and may not commute. The ordering of commuting local terms still matters. For a local Hamiltonian with $m$ terms, there are $m-1$ actions for each Trotter step. Any permutation on $m$ elements can be decomposed to a product of $\mathcal{O}(m^2)$ adjacent transpositions. Therefore our action set is universal, and a well-trained agent can rapidly steer the ordering to the optimum. The agent takes actions in sequence, the size of the action-space is $2^{n(m-1)}$. A deep neural network with $n(m-1)$ output neurons determines the action probabilities. We iteratively train the agent from measurement results on the output state $|\Psi_{\text{f}}\rangle$, the agent updates the path to maximize its total reward. Fig.~\ref{diagram} shows the diagram of our method.

	The reward of the agent received in each step is
	\begin{equation}
		R_{t}=
		\begin{cases}
			0, & t\in\{0,1,\dots,N_d-1\}\\
			\mathcal{R}, & t=N_d
		\end{cases}
		\label{eq:reward1}
	\end{equation}
	where $N_d$ is the time delay to get the reward, $\mathcal{R}$ is the modified reward function. In particular, we define
	$\mathcal{R}$ as
	
	\begin{equation}
		\mathcal{R}=\begin{cases}
			-1, &\text{ if } E\le E_{\mathrm{std}}\\
			-1/\log[\operatorname{clip}((E/E_{\mathrm{std}}-1),0.01,1.99)],  &\text{otherwise}
		\end{cases}
		\label{eq:reward2}
	\end{equation}
	where $E$ is the output energy given by our RL-steered path, $E_{\mathrm{std}}$ is the energy given by the standard repetition path without optimization. In order to avoid divergence of the reward, we use a $\operatorname{clip} $ function to clip the value of $(E/E_{\mathrm{std}}-1)$ within a range $(0.01,1.99)$.

	\section{Results}
	\subsection{Transverse-field Ising model}
	We first consider the one-dimensional transverse-field Ising model. With no assumption about the underlying structure, we initialize all qubits in the product state $|\Psi_{\mathrm{init}}\rangle = (|0\rangle + |1\rangle)^{\otimes N}/\sqrt{2^N}$. The Hamiltonian can be written as
	\begin{equation}
		\hat{H}^{\mathrm{TFI}} = -\sum_{j} (J \hat{Z}_{j}\hat{Z}_{j+1} + h \hat{X}_{j}).
	\end{equation}
	In the following, we choose $J = h = 1$. The system is in the gapless phase. For finite-size systems, the ground state of $\hat{H}^{\mathrm{TFI}}$ is $2$-UGS, therefore $2$-UDP and $2$-UDA.

	In the standard QITE, the ordering of local terms in each Trotter step is the same, e.g., we put commuting terms next to each other and repeat the ordering $\hat{X}_1, \dots, \hat{X}_N, \hat{Z}_1\hat{Z}_2, \dots, \hat{Z}_{N-1}\hat{Z}_{N}$. The quantum circuit of the standard QITE with $4$ qubits is shown in Fig.~\ref{fig:2}(a). Inspired by the randomization technique to speed up quantum simulation~\cite{childs2019faster, campbell2019random}, we also consider a randomized QITE scheme where we randomly shuffle the ordering in each Trotter step. There is no large quality difference between randomizations, and we pick a moderate one. In the RL-steered QITE, the reward is based on the expectation value of the output state $|\Psi_{\text{f}}\rangle$ on the target Hamiltonian 
	\begin{equation}
		E = \langle \Psi_{\text{f}} | \hat{H}^{\mathrm{TFI}} | \Psi_{\text{f}} \rangle.
	\end{equation}
	The lower the energy, the higher the reward. The RL agent updates the orderings step by step.
	
\begin{figure}[t]
		\centering
		\includegraphics[width=8.5cm]{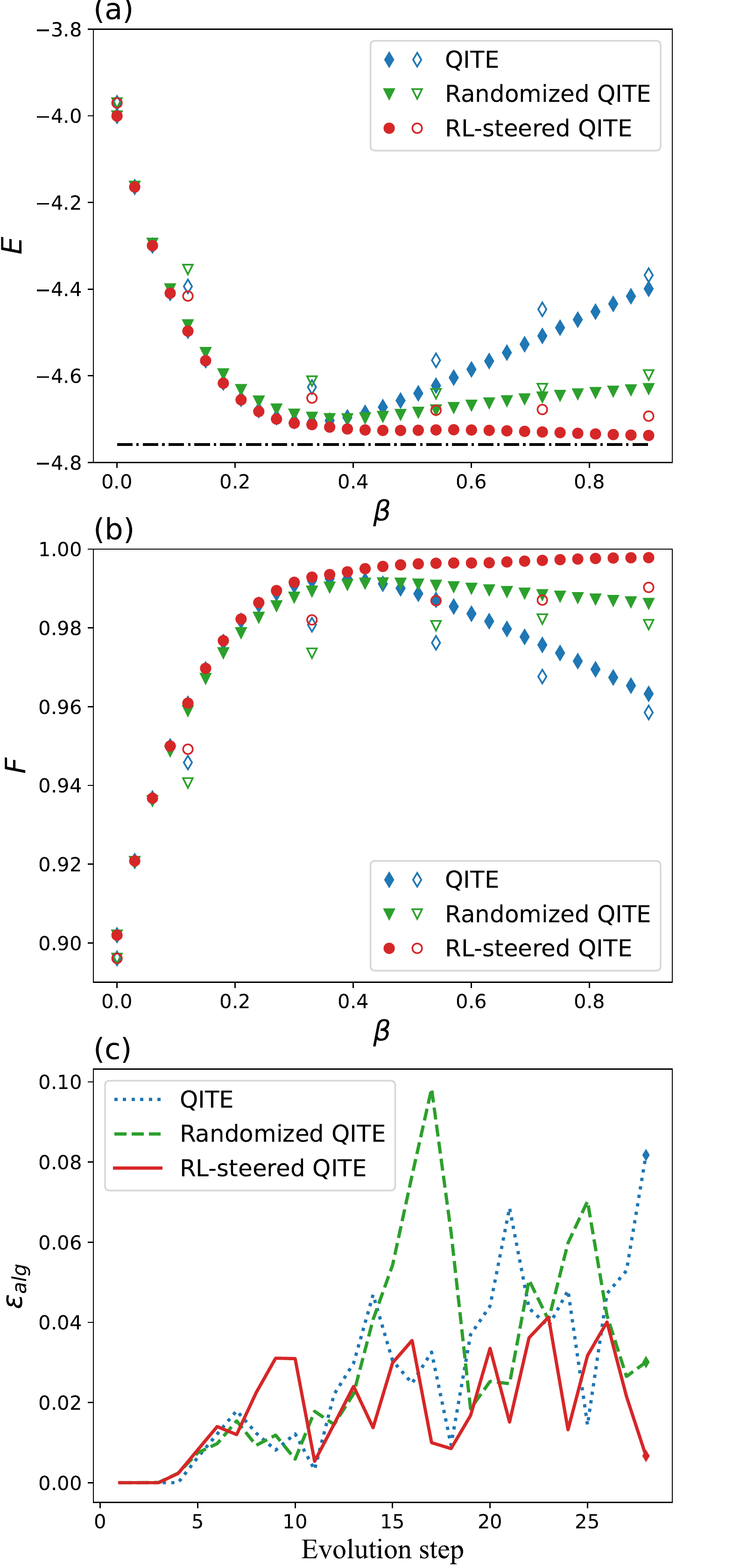}
		\caption{\textbf{Different QITE schemes for the transverse-field Ising model}. Filled markers represent the numerical data; unfilled markers represent the experimental data. (a) Energy versus $\beta$, the black dashed line represents the ground state energy.  (b) Fidelity versus $\beta$. (c) Algorithmic errors during the evolution.}
		\label{fig:tfim_results}
	\end{figure}

	For any given $\beta$, the RL agent can steer the QITE path and maximize the reward. We fix the system size $N = 4$, the number of Trotter steps $n = 4$, and the domain size $D = 2$. A numerical comparison of energy/fidelity obtained by the standard, the randomized and the RL-steered QITE schemes for different $\beta$ values is shown in Fig.~\ref{fig:tfim_results}(a)(b). The RL-steered path here is $\beta$-dependent. Throughout this paper we use the fidelity defined by $ F(\rho,\sigma)=\operatorname{Tr}\sqrt{\sigma^{1/2}\rho\sigma^{1/2}}$.

	When $\beta$ is small, RL cannot markedly decrease the energy since the total quantum resource is limited.  With the increase of $\beta$, the imaginary time evolution target state $ |\bar{\Psi}_{\text{f}}^{\prime}\rangle=e^{-\beta \hat{H}^{\mathrm{TFI}}}\left|\Psi_{\mathrm{init}}\right\rangle$ approaches the ground state, therefore the obtained energy of all paths decrease in the beginning. However, algorithmic errors increase with $\beta$, and this factor becomes dominant after a critical point, the energy of the standard/randomized QITE increases when $\beta > 1/3$. Accordingly, the fidelity increases first, then decreases. The RL-steered QITE outperforms the standard/randomized QITE for all $\beta$ values. Algorithmic errors in this path canceled out. The fidelity between the output state and the ground state constantly grows to $F > 0.996$. The gap between the ground state energy and the minimum achievable energy of the standard QITE is 0.053, and that of the RL-QITE is only 0.016. For a detailed optimized path see Appendix~\ref{sec3}.

Further, we implement the same unitary evolutions on a $4$-qubit liquid state nuclear magnetic resonance (NMR) quantum processor~\cite{nmrqc}.   We carry out the experiments with a $300$-MHz Bruker AVANCE III spectrometer. The processor is $^{13}$C-labelled trans-crotonic acid in a $7$-T magnetic field. The resonance frequency for $^{13}$C nuclei is about 75~MHz. The Hamiltonian of the system in a rotating frame is 

	\begin{equation}
		\hat{H}_{\text{S}}=\sum_{j=1}^{4} \pi \nu_{j} \hat{Z}_{j}+\sum_{1\leq i<j\leq 4}^{4} \frac{\pi}{2} J_{ij} \hat{Z}_{i} \hat{Z}_{j},
	\end{equation}
	where $\nu_j$ is the chemical shift of the $j$-th nuclei,  $J_{ij}$ is the J-coupling between the $i$-th and $j$-th nuclei, $\hat{Z}_{j}$ is the Pauli matrix $\sigma_z$ acting on the $j$-th nuclei. All the parameters can be found in Fig.~\ref{fig:2}(b). The quantum operations are realized by irradiating radiofrequency pulses on the system. We optimzie the pulses over the fluctuation of the chemical shifts of the nuclei with the technique of gradient ascent pulse engineering~\cite{grape}. The experiment is divided into three steps: (i) preparing the system in to a pseudo-pure state using the temporal average technique~\cite{pps}; (ii) applying the quantum operations; (iii) performing measurements~\cite{qst}.

	Denote the NMR output state as $\rho$, whose density matrix can be obtained through quantum state tomography. $\rho$ is a highly mixed state since quantum noise is inevitable.  We use the virtual distillation technique to
	suppress the noise~\cite{koczor2020exponential, huggins2020virtual}. The dominant eigenvector of $\rho$, $\lim_{M\rightarrow \infty}\rho^M/\text{Tr}(\rho^M)$, can be extracted numerically. Its expectation value on $\hat{H}^{\mathrm{TFI}}$ and its fidelity with the ground state are shown in Fig.~\ref{fig:tfim_results}(a)(b) with unfilled markers. Consistent with our numerical results, the RL-steered path significantly outperforms the other two for large $\beta$.

	In our simulation, we have four orderings to optimize and 28 local unitary operations to implement. Denote $|\Psi_{k}\rangle$ as the state after the $k$-th operation, $|\bar{\Psi}_{k}^{\prime}\rangle$ as the temporal target state with the ideal imaginary time evolution. The instantaneous algorithmic error during the evolution can be characterized by the squared Euclidean distance between $|\Psi_{k}\rangle$ and $|\bar{\Psi}_{k}^{\prime}\rangle$,
	\begin{equation}
		\epsilon_{\mathrm{alg}} = \||\Psi_{k}\rangle - |\bar{\Psi}_{k}^{\prime}\rangle\|^2.
	\end{equation}
	For $\beta=0.9$, Fig.~\ref{fig:tfim_results}(c) shows $ \epsilon_{\mathrm{alg}}$ as a function of evolution step $k$. Although $\epsilon_{\mathrm{alg}}$ fluctuates in all paths, it accumulates obviously in the standard QITE and eventually climbs to $\epsilon_{\mathrm{alg}}=0.082$. The randomized QITE performs slightly better and ends with $\epsilon_{\mathrm{alg}}=0.030$. The RL-steered QITE is optimal, the trend of $\epsilon_{\mathrm{alg}}$ shows no accumulation and drops to $\epsilon_{alg}=0.007$ in the end. Although we cannot directly estimate $\epsilon_{\mathrm{alg}}$ in experiments, we can minimize it via maximizing the reward function.
	
	\begin{figure}[t]
		\centering
		\includegraphics[width=8.5cm]{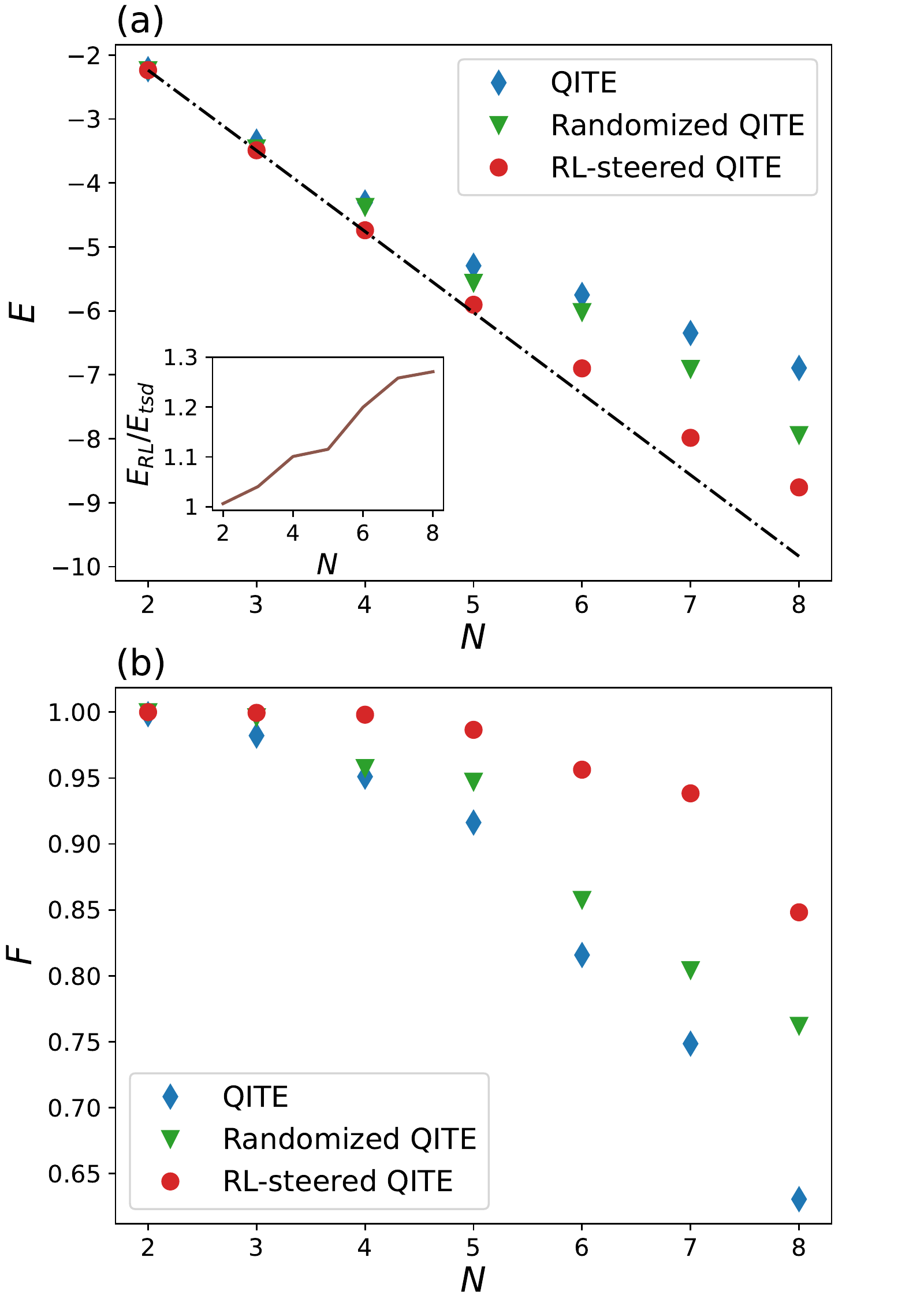}
		\caption{\textbf{Scaling of different QITE schemes.} Blue diamonds represent the standard quantum imaginary time evolution (QITE); green triangles represent the randomized QITE;  red circles represent the reinforcement learning (RL)-steered QITE. (a) Energy versus system size, and the energy ratio $E_{RL}/E_{\mathrm{std}}$ versus system size (inset). The black dash-dotted line represents the ground state energy. (b) Fidelity versus system size.}
		\label{fig:scaling_results}
	\end{figure}

	One question that arises is whether we can enhance the QITE algorithm with RL for larger systems? Now we apply our approach to the transverse-field Ising model with system sizes $N=2,3,\dots,8$ to demonstrate the scalability. Still, we consider the QITE with 4 Trotter steps. Denote the target state with ``evolution time" $\beta$ as
	\begin{equation}
		|\Psi'_{\text{f}}(\beta)\rangle = \frac{e^{-\beta \hat{H}^{\mathrm{TFI}}}}{\sqrt{\langle\Psi_{\mathrm{init}}|e^{-2\beta \hat{H}^{\mathrm{TFI}}}| \Psi_{\mathrm{init}}\rangle}}|\Psi_{\mathrm{init}}\rangle.
	\end{equation}
	In the following, we use an adaptive $\beta$ for different $N$ such that the expectation value $\langle\Psi'_{\text{f}}(\beta)|\hat{H}^{\mathrm{TFI}}|\Psi'_{\text{f}}(\beta)\rangle$ is always higher than the ground state energy of $\hat{H}^{\mathrm{TFI}}$ by $1\times 10^{-3}$. The results are illustrated in Fig.~\ref{fig:scaling_results}, the RL agent can efficiently decrease the energy and increase the fidelity between the final state and the ground state for all system sizes. The ratio of the RL-steered energy ($E_{\mathrm{RL}}$) to the standard QITE energy ($E_{\mathrm{std}}$) is also given. This ratio increases steadily with the number of qubits. Note that the neural networks we use here only contain four hidden layers. The hyperparameters were tuned for the $N=4$ case. We apply the same neural networks to larger $N$, the required number of training epochs does not increase obviously. If we want to increase the fidelity further for $N>4$, we can use more Trotter steps and tune the hyperparameters accordingly. There is little doubt, however, the training process will be more time-consuming. 
	
	\subsection{Sherrington-Kirkpatrick model}
	The second model we apply our method to is the Sherrington-Kirkpatrick (SK) model~\cite{sherrington1975solvable}, a spin-glass model with long-range frustrated ferromagnetic and antiferromagnetic couplings. Finding a ground state of the SK model is NP-hard~\cite{mezard1987spin}. On NISQ devices, solving the SK model can be regarded as a special Max-Cut problem and dealt with by the quantum approximate optimization algorithm~\cite{farhi2014quantum, farhi2019quantum}. Here we use the QITE to prepare the ground state of the SK model. Compared with the quantum approximate optimization algorithm, the QITE does not need to sample a bunch of initial points and implement classical optimization with exponentially small gradients~\cite{mcclean2018barren}. 
	
	Consider a six-vertex complete graph shown in Fig.~\ref{fig:2}(c). The SK model Hamiltonian can be written as
	
	\begin{equation}
		\hat{H}^{\mathrm{SK}} =\sum_{ i < j } J_{ij} \hat{Z}_{i}\hat{Z}_{j},
	\end{equation}
	we independently sample $J_{ij}$ are from a uniform distribution $J_{ij} \sim U(-1,1)$.

\begin{figure}[th]
		\centering
		\includegraphics[width=8.5cm]{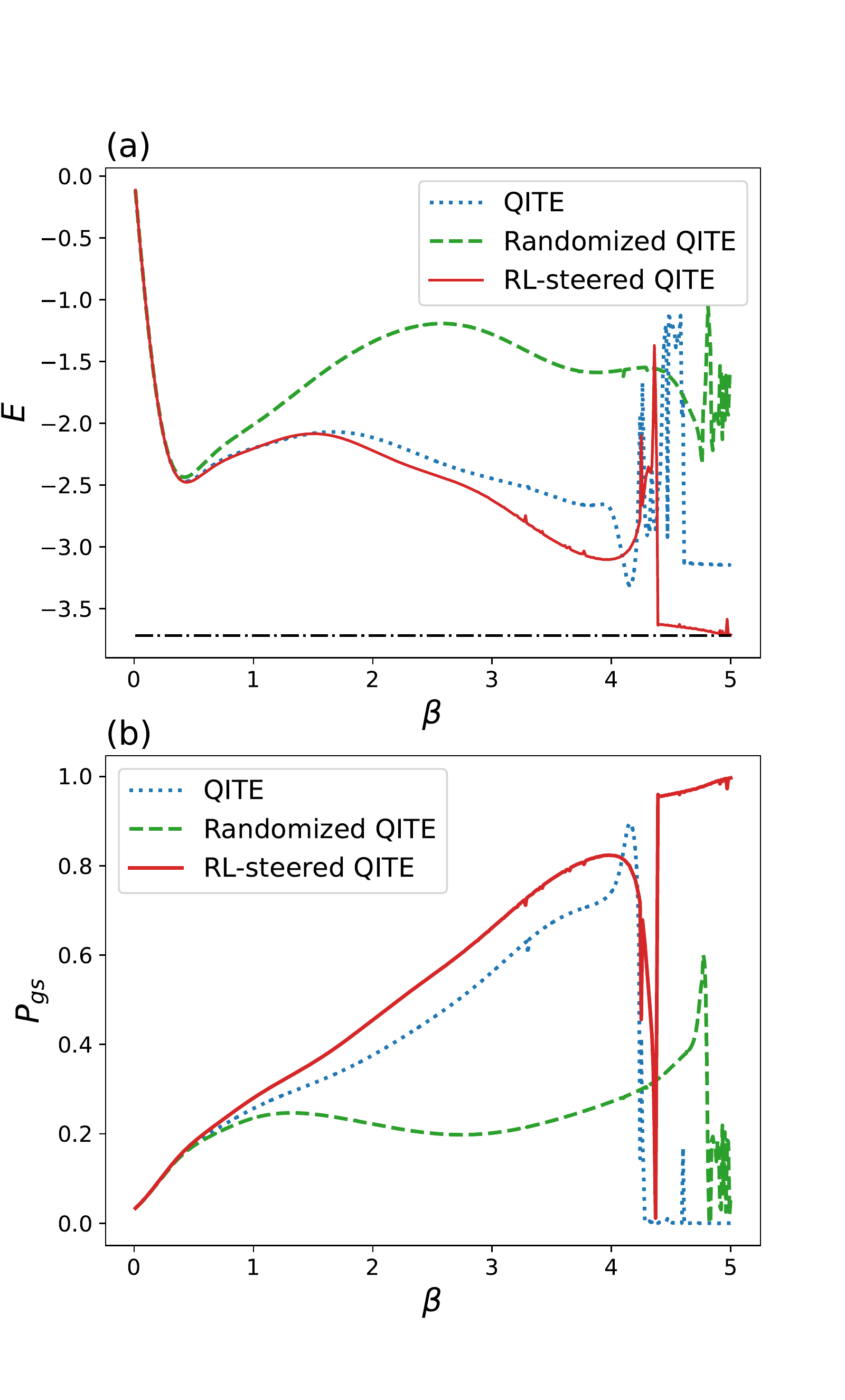}
		\caption{\textbf{Different QITE schemes for the Sherrington-Kirkpatrick model.} Blue dotted lines represent the standard quantum imaginary time evolution (QITE); green dashed lines represent the randomized QITE;  red lines represent the reinforcement learning (RL)-steered QITE. (a) Energy versus the imaginary time $\beta$. The black dash-dotted line represents the ground state energy. (b) Success probability versus the imaginary time $\beta$.}
		\label{fig:maxcut_results}
	\end{figure}

	Since $\hat{Z}\hat{Z}$-terms commute, there is no Trotter error in Eq.~\eqref{trotter} for the SK model. The ground state of $\hat{H}^{\mathrm{SK}}$ is two-fold degenerate. The QITE algorithm can prepare one of the ground states. We fix $\beta = 5$, $n = 6$, $D = 2$, sample $J_{ij}$ and train the agent to steer the QITE path. Define the probability of finding a ground state of $\hat{H}^{\mathrm{SK}}$ through measurements as $P_{\mathrm{gs}}$. Energy and $P_{\mathrm{gs}}$ as functions of $\beta$ are shown in Fig.~\ref{fig:maxcut_results}. Remember that the RL-steered path here was only optimized for a specific $\beta$ value (i.e., $\beta = 5$) since we want to verify the dependence of the ordering on $\beta$.

	For each path, we observe a sudden switch from a high probability of success to a low one when $4 < \beta < 5$. The reason is that for some states when the step interval exceeds a critical value, there is an explosion of LA error which utterly ruins QITE. After the critical value, the QITE algorithm loses its stability, and the energy $E$ fluctuates violently. The randomized QITE performs even worse than the standard one, where the highest success probability is only $0.60$. In the RL-steered QITE, $P_{\mathrm{gs}}$ falls down a deep ``gorge" but can recover soon. When $\beta = 5$, most algorithmic errors disappeared and $P_{\mathrm{gs}}= 0.9964$. In comparison, the standard QITE ends with $P_{\mathrm{gs}} = 0.0002$ and the randomized QITE ends with $P_{\mathrm{gs}} = 0.0568$, they dropped sharply and cannot be recovered even if we further improve $\beta$. For detailed couplings and QITE path see Appendix~\ref{sec3}.

	\section{Conclusions}
	We have proposed a RL-based framework to steer the QITE algorithm for preparing a $k$-UGS state. The RL agent can find the subtle evolution path to avoid error accumulation. We compared our method with the standard and the randomized QITE; numerical and experimental results demonstrate a clear improvement. The RL-designed path requires a smaller domain size and fewer Trotter steps to achieve satisfying performance for both the transverse-field Ising model and the SK model. We also noticed that randomization cannot enhance the QITE consistently, although it plays a positive role in quantum simulation. For the SK model, with the increase of the total imaginary time $\beta$, a switch from a high success probability to almost 0 exists. The accumulated error may ruin the QITE algorithm all of a sudden instead of gradually, which indicates the importance of an appropriate $\beta$ for high-dimensional systems. The RL-based method is a winning combination of machine learning and quantum computing. Even though we investigated only relatively small systems, the scheme can be directly extended to larger systems. The number of neurons in the output layer of the RL agent only grows linearly with system size $N$. A relevant problem worth considering is how to apply the QITE (or the RL-steered QITE) for preparing a quantum state that is $k$-UDP or $k$-UDA but not $k$-UGS.

	RL has a bright prospect in the NISQ era. In the future, one may use RL to enhance Trotterized/variational quantum simulation~\cite{sieberer2019digital, benedetti2020hardware, barison2021efficient, bolens2021reinforcement} similarly, but the reward function design will be more challenging. Near-term quantum computing and classical machine learning methods may benefit each other in many ways. Their interplay is worth studying further.

	
	\appendix
	\section{Local Approximation in QITE}\label{sec1}
	In the quantum imaginary time evolution algorithm, each non-unitary Trotter step is replaced by a unitary evolution. Our goal is to minimize the difference between $|\bar{\Psi}'\rangle$ and $e^{-i \Delta \tau \hat{A}[j]}|\Psi\rangle$. For a small step interval $\Delta \tau$, we have 
	\begin{equation}
		e^{-i \Delta \tau \hat{A}[j]}|\Psi\rangle \approx (1-i\Delta \tau \hat{A}[j])|\Psi\rangle.
	\end{equation}
	The modified goal is to minimize
	\begin{equation}
		\|\frac{|\bar{\Psi}^{\prime}\rangle-|\Psi\rangle}{\Delta \tau} +   i \hat{A}[l]|\Psi\rangle \|,
	\end{equation}
	where  $|\bar{\Psi}^{\prime}\rangle$ can be further approximated by
	\begin{equation}
		|\bar{\Psi}^{\prime}\rangle \approx \frac{(1-\Delta \tau \hat{h}[j])|\Psi\rangle}{\sqrt{1-2 \Delta \tau\langle\Psi|\hat{h}[j]| \Psi\rangle}}
	\end{equation}
	since $e^{- \Delta \tau \hat{h}[j]}|\Psi\rangle \approx (1-\Delta \tau \hat{h}[j])|\Psi\rangle$, $\langle\Psi|e^{-2 \Delta \tau \hat{h}[j]}| \Psi\rangle \approx 1-2 \Delta \tau\langle\Psi|h[j]| \Psi\rangle$.
	
	If we decompose the $D$-qubit operator $\hat{A}[j]$ in the Pauli basis,
	\begin{equation}
		\hat{A}[j]=\sum_{i_{1} \ldots i_{D}} a[j]_{i_{1} \ldots i_{D}} \hat{\sigma}_{i_{1}} \ldots \hat{\sigma}_{i_{D}},
	\end{equation}
	the minimum value of $\|\frac{|\bar{\Psi}^{\prime}\rangle-|\Psi\rangle}{\Delta \tau} +   i \hat{A}[l]|\Psi\rangle \|$ can be approximately achieved by solving the linear equation
	\begin{equation}
		\left(\mathbf{S}+\mathbf{S}^{T}\right) \mathbf{a}=-\mathbf{b}.
	\end{equation}
	$\mathbf{S}$ and $\mathbf{b}$ are some expectation values obtained by measurering $| \Psi\rangle$. Specifically,
	\begin{equation}
		S_{I J}=\langle\Psi|\hat{\sigma}_{I}^{\dagger} \hat{\sigma}_{J}| \Psi\rangle,
	\end{equation}
	\begin{equation}
		b_{I}=i\langle\Psi|\hat{\sigma}_{I}^{\dagger}| \Delta_{0}\rangle-i\left\langle\Delta_{0}\left|\hat{\sigma}_{I}\right| \Psi\right\rangle,
	\end{equation}
	where $\hat{\sigma}_{I}$, $\hat{\sigma}_{J}$ are Pauli operators, $\left|\Delta_{0}\right\rangle=(|\bar{\Psi}^{\prime}\rangle-|\Psi\rangle)/\Delta \tau$.
	
	\section{DPPO}\label{sec2}
	
	The DPPO algorithm combines the advantages of the policy-based RL and the value-based RL, aiming to avoid having too many policy updates. In DPPO, the actor builds a network to evaluate the policy $\pi_{\boldsymbol{\theta}}$ and takes actions according to the policy $\pi_{\boldsymbol{\theta}}$. The critic builds a network to evaluate the state
	value function $v_{\boldsymbol{\phi}}(s)$. The loss function of DPPO reads

	\begin{equation}
	\begin{aligned}
	L(\boldsymbol{\theta},\boldsymbol{\phi}) = &\hat{\mathbb{E}}_{\pi_{\boldsymbol{\theta}_{\text{old}}}}
	[ \min( r_{\boldsymbol{\theta}_{\text{old}}}(a|s_{0}, \boldsymbol{\theta})
	A_{\boldsymbol{\phi}}(s_{0},a),  \\&\operatorname{clip}
	\left(r_{\boldsymbol{\theta}_{\text{old}}}(a|s_{0}, \boldsymbol{\theta}), 1-\epsilon, 1+\epsilon \right)
	A_{\boldsymbol{\phi}}(s_{0},a))],  
	\end{aligned}
	\end{equation}

	where $\epsilon$ is the clipping parameter. The expectation $ \hat{\mathbb{E}}_{\pi_{\boldsymbol{\theta}_{\text{old}}}}$
	means we take empirical average over a finite batch of samples under the policy $\pi_{\boldsymbol{\theta}_{\text{old}}}$. $A$ is the advantage function  which measures the average state value of the action $a$ with respect to the state $s$ under the policy $\pi$. The term
	$r_{\boldsymbol{\theta}_{\text{odd}}}(a|s, \boldsymbol{\theta})$ is defined as the ratio of
	likelihoods
	\begin{equation}
		r_{\boldsymbol{\theta}_{\text{old}}}(a|s, \boldsymbol{\theta}) = \frac{\pi_{\boldsymbol{\theta}}(a | s)}{\pi_{\boldsymbol{\theta}_{\text{old}}}(a |s)}.
	\end{equation}
	The clip function with $c \le d$ is defined by
	\begin{equation}
		\label{eq:6}
		\text{clip}(f(x), c, d) =
		\begin{cases}
			d, & \text{ if } f(x) > d, \\
			f(x), & \text{ if } c \le f(x) \le d, \\
			c, & \text{ if } f(x) < c.
		\end{cases}
	\end{equation}
	$\operatorname{clip} \left(r_{\boldsymbol{\theta}_{\text{old}}}(a|s_{0}, \boldsymbol{\theta}), 1-\epsilon, 1+\epsilon \right)$ penalizes large changes between nearest updates, which corresponds to the trust region of the first order policy gradient. The loss function poses a lower bound on the improvement induced by an update and hence establishes a trust region around $\pi_{\boldsymbol{\theta}_{\text {old}}}$. The hyperparameter $\epsilon$ controls the maximal improvement and thus the size of the trust region.

	In our computation, the RL agent uses one deep neural network to approximate both the value of the state and the value of the policy. The neural network consists of four hidden layers. All layers have ReLU activation functions except the output layer. Appendix Table 1 summarizes the hyper-parameters of the neural network.
	
	\begin{table}
		\begin{center}
			\small{Appendix Table 1: Training Hyper-Parameters in DPPO}
		\end{center}
		\centering
		\begin{tabular}{cc}
			\textrm{Hyperparameters} &\textrm{Values} \\
			\hline
			Optimizer & Adam\\
			Neurons in neural network & $\{1024,1024,1024,1024\}$\\
			Evaluator number & 12\\
			DPPO clipping $\epsilon$ & 0.2\\
			Learning rate & $0.0003$\\
			Update steps & 4\\
			Reward decay $\gamma$ & 0.99\\
			
			\label{para}
		\end{tabular}
	\end{table}

	DPPO consists of multiple independent agents (evaluators) with their weights, who interact with a different copy of the environment in parallel. The data collected by the evaluators are assumed to be independent and identically distributed. Further, they can explore a more significant part of the state-action space in much less time.
	
	We modify the original DPPO algorithm for better performance. For the critic, we normalize the advantage so that the value estimation is stable. The $\operatorname{clip} $ function is used in both the policy loss and the value loss. For the actor, instead of using the traditional $\epsilon$-greedy method~\cite{sutton} to sample the policy, we use the entropy regularization method, which was first introduced in the soft actor-critic algorithm~\cite{SAC}. 
	
	\begin{figure}[t]
		\centering
		\includegraphics[width=8.5cm]{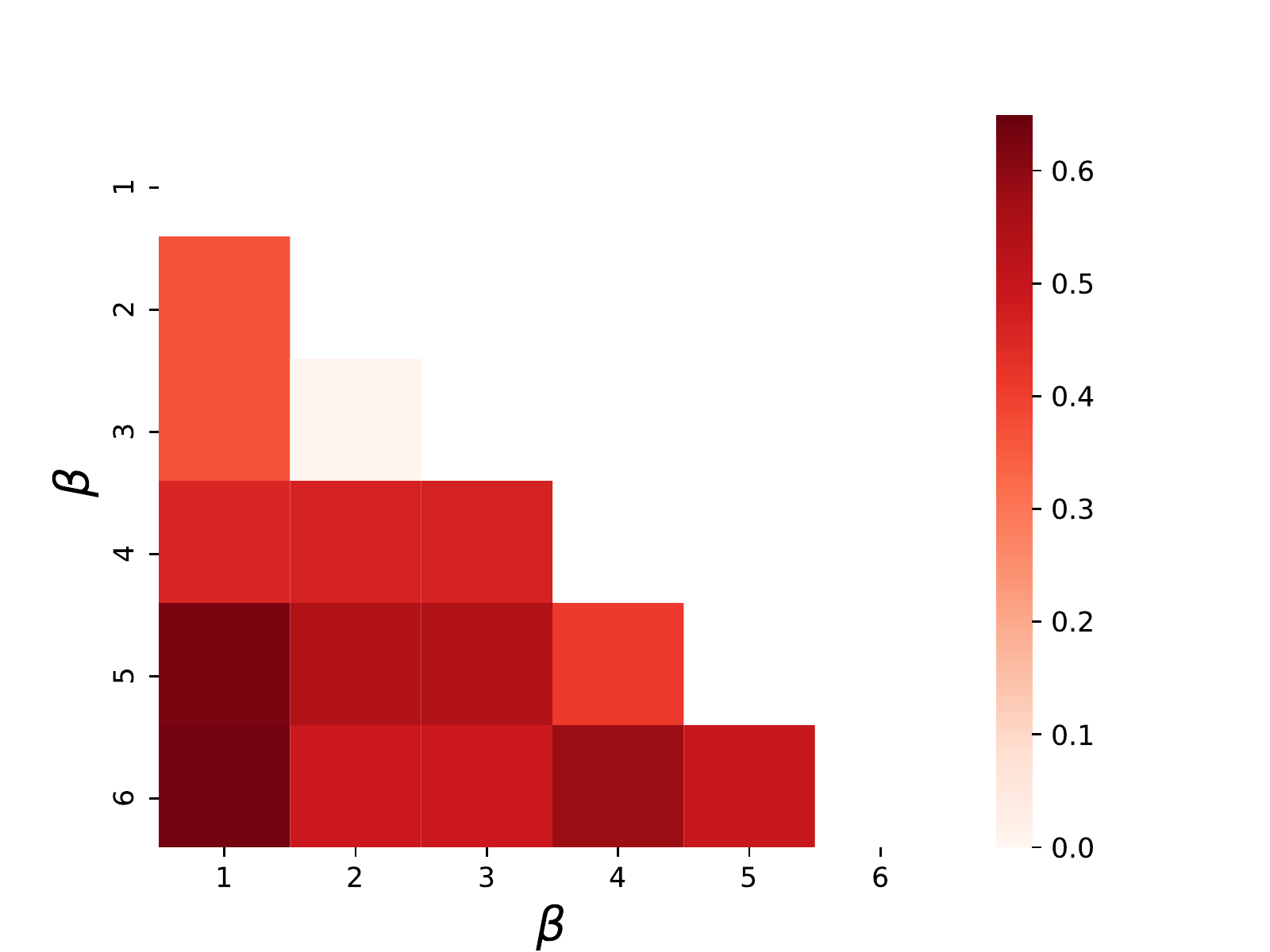}
		\caption{The Hamming distance matrix for the RL agent minima protocols.}
		\label{fig:hamming}
	\end{figure}  
	
	For the SK model, we found that the control landscape undergoes a phase-transition-like behavior, as reported in the quantum state preparation paradigms~\cite{bukov2018reinforcement,PhysRevA.97.052114,Larocca_2018,PhysRevLett.122.020601}. We compare the policies trained with different $\beta$ values to understand this transition better. The Hamming distance matrix is shown in Fig.~\ref{fig:hamming}.  For small $\beta$ values, the energy landscape is almost convex, and we can easily find the unique optimal protocol. With the increase of $\beta$, the barriers between local minimums get higher. When $\beta > 4$, the control landscape has many local minimums separated by extensive walls.
	
\begin{figure}[t]
	\centering
	\includegraphics[width=8.5cm]{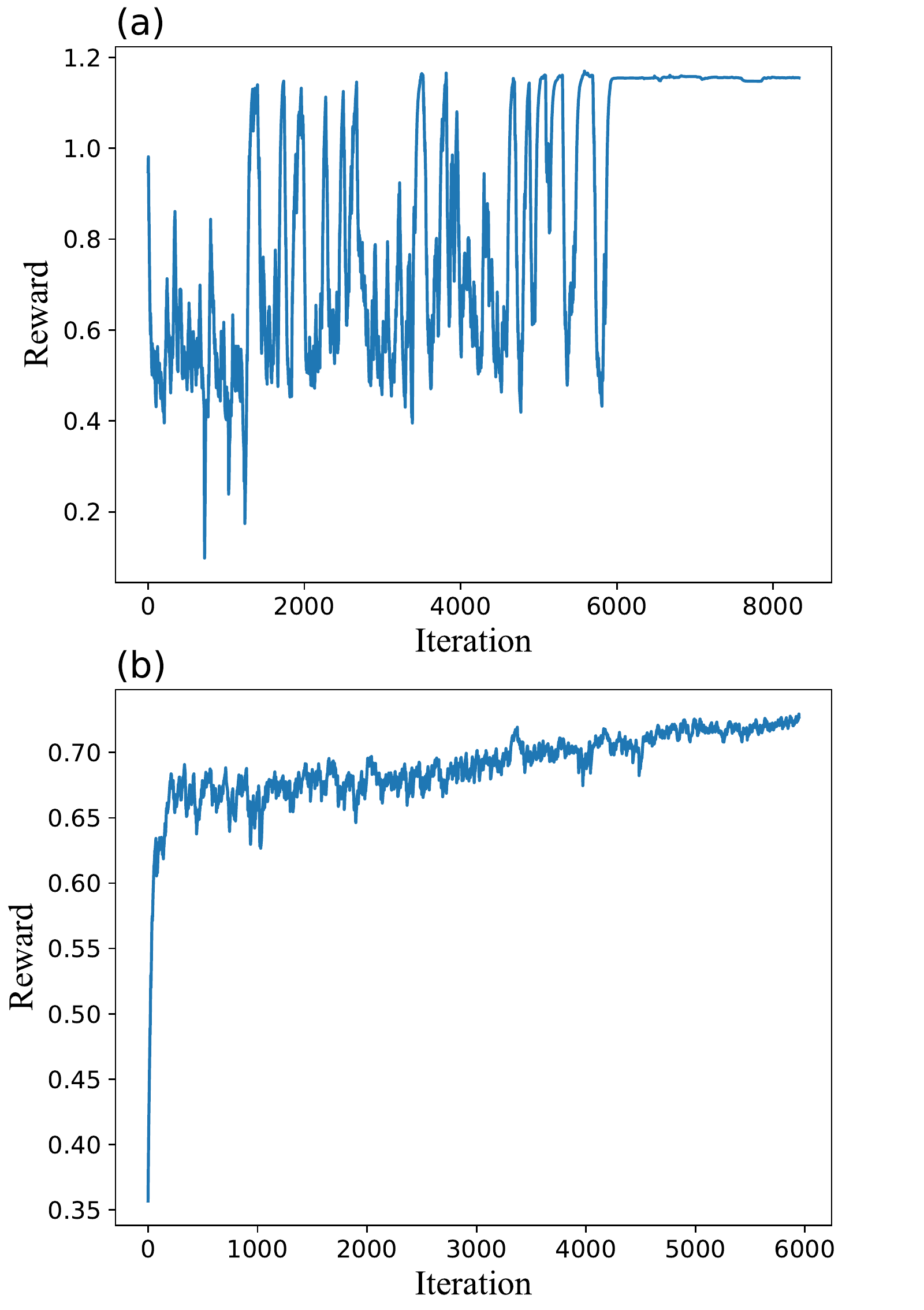}
	\caption{Learning curves of RL-steered QITE. (a) The transverse-field Ising model with $\beta=0.9$. (b) The SK model with $\beta=5$.}
	\label{fig:lr_curve}
\end{figure}

	Numerical codes were written with Python 3.8 and PyTorch 1.8~\cite{paszke2019pytorch}. Numerical calculations were run on two 14-core 2.60GHz CPUs with 188 GB memory and four GPUs. We plot the RL learning curves in Fig.~\ref{fig:lr_curve}. The reward converges after about 6000 iterations.

\section{Optimized Paths}\label{sec3}
This section shows two detailed paths designed by the RL agent. Although the RL agent can find efficient paths by updating the ordering, they are challenging to interpret. The path for the transverse-field Ising model is given in Appendix Table 2. The sampled couplings of the SK model are given in Appendix Table 3, and the corresponding path is shown in Appendix Table 4.

\begin{table}[H]\label{TFIM_ordering}
	\begin{center}
		\small{Appendix Table 2: RL-optimized path for the transverse-field Ising model with $N = 4, \beta = 0.9$}
	\end{center}
	\centering
	\begin{tabular}{ccccccccc}
		\rowcolor[gray]{.9}
		Trotter Step&&&&&&&\\
		\rowcolor[gray]{.85}
		1&$X_4$&$X_3$&$X_2$&$Z_1Z_2$&$Z_2Z_3$&$Z_3Z_4$&$X_1$\\
		\rowcolor[gray]{.8}
		2&$Z_3Z_4$&$X_3$&$X_2$&$X_4$&$Z_2Z_3$&$Z_1Z_2$&$X_1$\\
		\rowcolor[gray]{.75}
		3&$Z_3Z_4$&$X_3$&$X_4$&$X_2$&$Z_2Z_3$&$Z_1Z_2$&$X_1$\\
		\rowcolor[gray]{.7}
		4&$Z_3Z_4$&$X_3$&$X_4$&$Z_1Z_2$&$Z_2Z_3$&$X_1$&$X_2$\\
	\end{tabular}
\end{table}

\begin{table}[H]\label{SK_couplings}
	\begin{center}
		\small{Appendix Table 3: Sampled couplings of the six-qubit SK model}
	\end{center}
	\centering
	\begin{tabular}{ccccccc}
		\rowcolor[gray]{.9}
		\diagbox{i}{$J_{ij}$}{j}&2&3&4&5&6\\
		\rowcolor[gray]{.85}
		1&0.5554&-0.5249&0.6465&0.9315&0.9452\\
		\rowcolor[gray]{.8}
		2&&-0.0931&0.2181&0.5511&0.2832\\
		\rowcolor[gray]{.75}
		3&&&0.4440&-0.9299&-0.4031\\
		\rowcolor[gray]{.7}
		4&&&&-0.8830&0.7141\\
		\rowcolor[gray]{.65}
		5&&&&&-0.2543\\
	\end{tabular}
\end{table}

\begin{table}[H]\label{SK_ordering}
	\begin{center}
		\small{Appendix Table 4: Sampled couplings of the six-qubit SK model}
	\end{center}
	\centering
	\begin{tabular}{cccccccccccccccccc}
		\rowcolor[gray]{.9}
		Trotter Step&&&&&&&&&&&&&&&\\
		\rowcolor[gray]{.85}
		1&$Z_4Z_5$&$Z_1Z_4$&$Z_1Z_5$&$Z_1Z_6$&$Z_1Z_2$&$Z_2Z_4$&$Z_2Z_5$&$Z_2Z_6$\\\rowcolor[gray]{.85}&$Z_2Z_3$&$Z_3Z_4$&$Z_3Z_6$&$Z_3Z_5$&$Z_4Z_6$&$Z_5Z_6$&$Z_1Z_3$&$ $
		\\
		\rowcolor[gray]{.8}
		2&$Z_1Z_2$&$Z_1Z_4$&$Z_1Z_3$&$Z_1Z_5$&$Z_1Z_6$&$Z_2Z_3$&$Z_2Z_4$&$Z_2Z_6$\\\rowcolor[gray]{.8}&$Z_3Z_4$&$Z_3Z_5$&$Z_2Z_5$&$Z_4Z_5$&$Z_4Z_6$&$Z_5Z_6$&$Z_3Z_6$&$ $
		\\
		\rowcolor[gray]{.75}
		3&$Z_1Z_2$&$Z_1Z_3$&$Z_1Z_4$&$Z_1Z_5$&$Z_1Z_6$&$Z_2Z_3$&$Z_2Z_4$&$Z_2Z_5$\\\rowcolor[gray]{.75}&$Z_2Z_6$&$Z_3Z_5$&$Z_3Z_6$&$Z_3Z_4$&$Z_4Z_6$&$Z_5Z_6$&$Z_4Z_5$&$ $
		\\
		\rowcolor[gray]{.7}
		4&$Z_1Z_3$&$Z_1Z_4$&$Z_1Z_2$&$Z_1Z_5$&$Z_1Z_6$&$Z_2Z_3$&$Z_2Z_5$&$Z_2Z_4$\\\rowcolor[gray]{.7}&$Z_2Z_6$&$Z_3Z_5$&$Z_3Z_4$&$Z_4Z_5$&$Z_3Z_6$&$Z_5Z_6$&$Z_4Z_6$&$ $
		\\
		\rowcolor[gray]{.65}
		5&$Z_1Z_3$&$Z_1Z_2$&$Z_1Z_4$&$Z_1Z_6$&$Z_2Z_3$&$Z_1Z_5$&$Z_2Z_4$&$Z_2Z_5$\\\rowcolor[gray]{.65}&$Z_2Z_6$&$Z_3Z_5$&$Z_3Z_4$&$Z_3Z_6$&$Z_4Z_6$&$Z_5Z_6$&$Z_4Z_5$&$ $
		\\
		\rowcolor[gray]{.6}
		6&$Z_1Z_2$&$Z_1Z_3$&$Z_1Z_5$&$Z_1Z_4$&$Z_1Z_6$&$Z_2Z_3$&$Z_2Z_5$&$Z_2Z_4$\\\rowcolor[gray]{.6}&$Z_3Z_4$&$Z_2Z_6$&$Z_3Z_5$&$Z_3Z_6$&$Z_4Z_5$&$Z_4Z_6$&$Z_5Z_6$&$ $
		\\

	\end{tabular}
\end{table}

	\bibliography{ref}

\end{document}